\documentstyle[preprint,aps,prd]{revtex}

\begin{document}

\title {Universal Approach to Cosmological Singularities in
        Two Dimensional Dilaton Gravity}

\author{Gabriel N. Martin$^1$ and Francisco D. Mazzitelli$^{1,2}$ }

\address{{\it
$^1$ Departamento de F\'\i sica, Facultad
de Ciencias Exactas y Naturales\\
Universidad de Buenos Aires- Ciudad Universitaria, Pabell\' on I\\
1428 Buenos Aires, Argentina\\
$^2$Instituto de Astronom\'\i a y F\'\i sica del Espacio\\
Casilla de Correo 67 - Sucursal 28\\
1428 Buenos Aires, Argentina}}

\maketitle

\begin{abstract}
We show that in a large class of two dimensional models
with conformal matter fields,
the semiclassical cosmological solutions have a
weak coupling singularity
if the classical matter content is below a certain
threshold. This threshold and the
approach to the singularity
are model-independent.
When the matter fields are not conformally invariant,
the singularity persists if the quantum state is
the vacuum near the singularity, and could dissappear
for other quantum states.

\end{abstract}

\vskip 3cm
\centerline{April 1994}

\newpage

\section{INTRODUCTION}

In the last two years there has been a lot of activity in
the study of (1+1)-dimensional models of gravity. The models
include the spacetime metric, a dilaton and $N$
conformal matter fields, and are useful toy models to
address the problem of black hole formation and evaporation,
including backreaction effects.

The original theory proposed by Callan, Giddings,
Harvey and Strominger (CGHS) is \cite{cghs}
\begin{eqnarray}
S_{CGHS} &=& {1\over 2\pi}\int d^2x\sqrt{-g}\left[e^{-2\phi}
(R+4(\nabla\phi)^2+4\lambda^2)-{1\over 2}\sum_{i=1}^N
(\nabla f_i)^2\right ]
\nonumber\\
&-&{\kappa\over 8\pi}\int d^2x\sqrt{-g} R{1\over\nabla^2}R
\label{Scghs}
\end{eqnarray}
where $\phi$ is the dilaton, $f_i$ are the matter fields
and $\kappa=N/12$. The term proportional to $\kappa$
is the well known conformal anomaly, and comes from
one-loop quantum corrections. The model
is exactly soluble at the classical level ($\kappa=0$).
However, the semiclassical CGHS equations have not been
analytically solved.

It is possible to modify the gravity-dilaton couplings of the
theory in order to find exactly soluble semiclassical equations
\cite{bil-da}-\cite{rst}. In particular, in Ref.\cite{rst},
Russo, Susskind and Thorlacius (RST) obtained a soluble model
imposing the classical conservation law
\begin{equation}
\partial_{\mu}\partial^{\mu}(\rho-\phi)=0\label{sym}
\end{equation}
to be preserved at the semiclassical level. This leads to
the RST semiclassical action
\begin{equation}
S_{RST}=S_{CGHS}-{\kappa\over 4\pi}\int d^2x\sqrt{-g}\phi R
\label{Srst}
\end{equation}
For $\kappa=(N-24)/12$ Eq.\ref{Srst} describes a conformal
field theory with vanishing total central charge.

The RST model is not only useful for the analysis of black
hole evaporation.
It has also been used to study the effect of backreaction on
the classical cosmological solutions.   In particular, it has
been shown  that, if the classical matter content is
below certain threshold, the semiclassical cosmological solutions
develop
a weak coupling singularity \cite{mr}. This result is
in contradiction
with the standard lore:
quantum effects do not help to smear the cosmological singularities
but generate new singularities at weak coupling. \footnote {A
result of this type has also been found
in Ref.\cite{klim} for $N=24$ in
a different model.}

Since the RST model contains an  additional symmetry
that leads to the conservation law Eq. \ref{sym},
one  may suspect that some of the results may not be generic,
and that other models like CGHS may give qualitatively
different results \cite{pistro}.
In this work we will
analyze the issue of the weak
coupling cosmological singularities in a model with
arbitrary graviton-dilaton couplings
\begin{eqnarray}
S &=& {1\over 2\pi}\int d^2x\sqrt{-g}\left[A(\phi)
(R+4(\nabla\phi)^2+4\lambda^2(\phi))-{1\over 2}\sum_{i=1}^N
(\nabla f_i)^2\right ]
\nonumber\\
&-&{\kappa\over 8\pi}\int d^2x\sqrt{-g} R{1\over\nabla^2}R
\label{S}
\end{eqnarray}
We will see that, for a large class of couplings $A(\phi)$
and $\lambda(\phi)$, the weak coupling singularity is
present. Moreover, we will prove that the threshold and the
leading behaviour of the scale factor near the singularity
are independent of the couplings.
Finally, we will consider  non-conformally invariant
matter fields. We will prove that for a particular choice
of the quantum state of the matter fields the
singularity  does not dissappear. However, we will also argue that,
for other
quantum states,
cosmological particle creation
may wash the
singularity away.

\section{COSMOLOGICAL SINGULARITIES}

In the conformal gauge $g_{++}=g_{--}=0, g_{+-}=-{1\over 2}e^{2\rho}$,
the equations of motion derived from Eq.\ref{S} are
\begin{eqnarray}
&&-A'\partial_+\partial_-\rho-4A \partial_+\partial_-\phi-2A'\partial_-
\phi\partial_+\phi-e^{2\rho}(\lambda\lambda 'A + {1\over 2}\lambda^2A')
=0\nonumber\\
&& A''\partial_+\phi\partial_-\phi +\kappa\partial_+\partial_-\rho
+A'\partial_+\partial_-\phi+\lambda^2Ae^{2\rho}=0
\nonumber\\
&&(A''-4A)(\partial_{\pm}\phi)^2+A'\partial_{\pm}\partial_{\pm}\phi
-2A'\partial_{\pm}\phi\partial_{\pm}\rho+{1\over 2}
\sum_{i=1}^N(\partial_{\pm} f_i)^2
\nonumber\\
&&-\kappa[(\partial_{\pm}\rho)^2-\partial_{\pm}\partial_{\pm}\rho]=
\kappa t_{\pm}\label{eqclas}
\end{eqnarray}
where the prime denotes a derivative with respect to $\phi$.
The functions $t_{\pm}(x^{\pm})$ depend on the quantum state
of the matter fields and come from the variation of the anomaly
term.

The solutions to the above equations can be trusted only in
{\it weak coupling
regions} where the quantum corrections to the semiclassical
action Eq.\ref{S} are small. These quantum corrections
will be supressed by a $\phi$-dependent coupling
given by $g_c^2\sim {1\over A'^2}$. If the quantum corrections
are calculated with the action Eq.\ref{S} that includes
the trace anomaly, they will be supressed by an effective
coupling $g_{eff}$ which is given
by the inverse of the determinant of the $\rho-\phi$ target space
metric in Eq.\ref{S} (see Refs.\cite{rt},\cite{mr}). For our
model we obtain
\begin{equation}
g_{eff}^2={1\over\vert A'^2-4\kappa A\vert}\label{geff}
\end{equation}
so the weak coupling region is defined by $g_{eff}\ll 1$
or $g_c\ll 1$.
For simplicitly in what follows we will assume that
$A>0$ and $A'^2\gg \kappa A$ in the weak coupling
region.

The equations of motion
can be written in a simpler form in terms of
the new fields $X$ and $Y$ defined as
\begin{eqnarray}
X &=& \kappa\rho + A(\phi)\nonumber\\
Y &=& \int d\phi\sqrt{\vert {1\over 4\kappa}A'^2-A\vert }
\label{xy}
\end{eqnarray}
Indeed, the equations of motion read
\begin{eqnarray}
\partial_+\partial_-X &=& - \lambda^2Ae^{2\rho}\nonumber\\
\partial_+\partial_-Y &=& {1\over 8Y'}e^{2\rho}
[A'\lambda^2+2\lambda\lambda'
A-{2\over\kappa}AA'\lambda^2]\nonumber\\
\kappa t_{\pm} &=& {1\over 2}\sum_{i=1}^N
\partial_{\pm}f_i\partial_{\pm}f_i
+ 4 \partial_{\pm}Y\partial_{\pm}Y
-{1\over \kappa}\partial_{\pm}X\partial_{\pm}X+\partial^2_{\pm}X
\label{xy-eq}
\end{eqnarray}
For $\lambda(\phi)=0$, $X$ and $Y$ are free fields and the
equations can be trivially solved. This fact will be
important in what follows.

Let us now consider time-dependent cosmological solutions.
In coordinates $\sigma^{\pm}=\tau\pm\sigma$ the two dimensional
metric is given by
\begin{equation}
ds^2=-e^{2\rho(\tau)}d\sigma^+d\sigma^-
\end{equation}
To solve the equations of motion we must fix the functions
$t_{\pm}$. Since the quantum matter fields are
conformally invariant, a natural choice for the quantum state
is the conformal vacuum\cite{bd}, in which $t_{\pm}(\sigma^{\pm})
=0$.
With this choice, the solutions for $\lambda=0$ are
\begin{eqnarray}
X&=&k_1\tau + const\\
Y&=&k_2\tau + const\\
0&=&{1\over 8}\sum_{i=1}^N
\dot f_i^2 -{1\over 4\kappa}k_1^2 + k_2^2\label{sol}
\end{eqnarray}
where $k_1$ and $k_2$ are integration constants.
If $\sum_{i=1}^N\dot f_i^2\neq 0$, $k_1$ cannot vanish and without loss
of generality we can choose the coordinate $\tau$ such that
$k_1={\kappa\over T}$, where $T$ is an arbitrary time scale.

{}From the above solution we observe that
the classical matter content  defined as   $m^2={1\over 8}
\sum_{i=1}^N\dot f_i^2 T^2$ must
satisfy $0\leq m^2\leq m^2_{cr}={\kappa\over 4}$.
If $m^2$ exceeds the critical value $m^2_{cr}$,
there is no solution.
When
$m^2$ equals the critical value,
$\phi$ is constant and $\rho$ is a linear function of $\tau$.
The spacetime metric describes  a two dimensional Milne
universe\cite{milne}.
The scalar curvature $R=2\ddot\rho e^{-2\rho}$ vanishes.
This is to be expected, since the Milne universe is
merely a
non trivial coordinatization of flat spacetime. This
result is independent of the function $A(\phi)$.

We will now show that, if $m^2<m^2_{cr}$, the semiclassical
solutions develop a
weak coupling singularity for $\tau\rightarrow -\infty$.
Let us denote
by $\delta$ the distance between the actual matter content
and the threshold, i.e., $\delta\equiv\sqrt{m^2_{cr}-m^2}$.
We obtain
\begin{eqnarray}
X&=&A(\phi)+\kappa\rho={\kappa\tau\over T}+const\nonumber\\
Y&=&\int d\phi\sqrt{{1\over 4\kappa}A'^2-A}=
{\delta\vert\tau\vert\over T}
+const
\label{sol2}
\end{eqnarray}
In the weak  coupling region we have $A'^2\gg \kappa A$. Therefore,
\begin{equation}
A-2\kappa\int d\phi{A\over A'}\approx 2\sqrt\kappa\delta
{\vert\tau\vert\over T}+const
\end{equation}
{}From this equation we see that in general
$A(\tau)$ {\it is not}
a linear function
\footnote
{In the particular case $A(\phi)$ proportional to $\phi^2$,
$Y$ is proportional to $A$. Therefore $A$ is linear in $\tau$
and we have a Milne universe for any matter content. However,
this particular $A(\phi)$ does not satisfy our hypothesis
$A'^2\gg \kappa A$}
of $\tau$. Consequently, $\ddot\rho =-{1\over\kappa}\ddot A$
is different from zero
and the scalar curvature $R$ diverges for $\tau\rightarrow -\infty$.

It is instructive to see the solutions in some particular
cases. The
condition  $A'^2\gg \kappa A$ is satisfied, for example, for
couplings of the form $A(\phi)=e^{-\gamma\phi}$, for
$\phi\rightarrow -\infty$. In this case
\begin{eqnarray}
& &\phi(\tau)\approx -{1\over\gamma}\ln [2\sqrt\kappa\delta{\vert
\tau\vert\over T}]\rightarrow -\infty\,\,\,(\tau
\rightarrow -\infty )\nonumber\\
& &A(\tau)\approx 2\sqrt\kappa\delta{\vert\tau\vert\over T}
+{2\kappa\over\gamma^2}
\ln [2\sqrt\kappa\delta{\vert
\tau\vert\over T}]\nonumber\\
& &\rho(\tau)\approx(1+{2\delta\over\sqrt\kappa}){\tau\over T}
- {2\over\gamma^2}
\ln [2\sqrt\kappa\delta{\vert
\tau\vert\over T}]\nonumber\\
%\end{eqnarray}
& & R(\tau)\approx{4\over\gamma^2\tau^2 }e^{-2\rho}\rightarrow\infty
\,\,\, (\tau\rightarrow -\infty)\label{examp}
\end{eqnarray}
Similar results can be obtained for other families of couplings.
For example, for $A(\phi)=\phi ^{2n},\,\,\,n>1$ we have
\begin{eqnarray}
& &\phi(\tau)\approx (2\sqrt\kappa\delta{\vert\tau\vert\over T}
)^{1\over 2n}
 \nonumber\\
& &A(\tau)\approx 2\sqrt\kappa\delta{\vert\tau\vert\over T}+
{\kappa\over n}(2\sqrt\kappa\delta{\vert\tau\vert\over T})^{1\over n}
\nonumber\\
& &\rho(\tau)\approx (1+{2\delta\over\sqrt \kappa}){\tau\over T}
-{1\over n}(2\sqrt\kappa\delta{\vert\tau\vert\over T})^{1\over n}
\nonumber\\
%\end{eqnarray}
& & R(\tau)\approx {2(n-1)\over n^3}
(2\sqrt\kappa\delta\vert\tau\vert)^{{1\over n}}
{1\over \tau^2} e^{-2\rho}\rightarrow\infty
\,\,\, (\tau\rightarrow -\infty)\label{examp2}
\end{eqnarray}

The general result is that, as long as  $A'^2\gg \kappa A$,
a weak coupling
singularity takes place. The threshold and the leading behaviour
of the Liouville field $\rho(\tau)$  are independent of the
function $A(\phi)$. Other quantities like the
subleading correction to  $\rho(\tau)$  and the
scalar curvature  do
depend on this function.

Up to here we assumed that the `potential' $\lambda^2$
vanishes. It is easy to see that
the results are valid for a large class of potentials.
Indeed, as long as
\begin{equation}
 R\gg max\{\lambda^2 A,\,\,\,
\vert(\lambda^2+2{A\over\dot A}\lambda\dot\lambda)
\vert\}\label{cond}
\end{equation}
the terms containing $\lambda^2$ can be neglected
in the equations of motion. As a  consequence,
the analysis of the existence of the singularities
needs no modifications. It is worth remarking that
the condition Eq.\ref{cond} is not too restrictive: the
scalar curvature diverges like
$e^{2(1+{2\delta\over\sqrt\kappa})\vert
\tau\vert}$ times a power of $\tau$
while $A(\tau)$ diverges linearly in $\tau$.
Therefore, the above
conditions are satisfied unless $\lambda (\phi)$ has a strong
divergence in the weak coupling region.

Finally we point out that, when $\lambda^2\neq 0$, we
cannot conclude from our calculations that there is no
solution when the  matter content is above the threshold.
Indeed, in the
RST model there is a solution for any matter
content \cite {mr}. Of course, the solution does not have weak coupling
singularities when
$m^2\geq m^2_{cr}$, in agreement with the analysis presented here.

The existence of the threshold and a similar dependence
with the logarithm of $\delta$ (see Eq.\ref{examp}) have
also been found
in the analysis of gravitational
collapse within the RST model \cite{strtor}.
Therefore our analysis suggests that, also in that problem,
the same
threshold  should appear for other theories. However, the
presence of $\log [\delta]$
in the expression for the mass of the black hole
would be particular of the exponential
couplings.

\bigskip

\section{NON MINIMAL COUPLING: BREAKING OF CONFORMAL INVARIANCE}

Let us now consider the case of non-conformally coupled
matter fields. We add to the classical action
the term
\begin{equation}
\Delta S={\xi\over 4\pi}\int d^2x\sqrt{-g}R\sum_{i=1}^N f_i^2
\label{delta-s}
\end{equation}
where $\xi$ is an arbitrary  constant. In principle, one
can compute the effective action and the effective semiclassical
equations induced by the  non conformal matter fields using an
expansion in powers of $\xi$. However, these equations are
non-local (even in the conformal gauge) and extremely difficult
to solve. We will follow here a simpler and
more qualitative approach.
It is well known that breaking of conformal invariance induces
particle creation.
The energy density of the created particles
(which we will denote by
$\epsilon$) contributes as a
classical source, i.e. with a positive sign, in Eq. \ref{sol}.
Therefore, if  $\epsilon\sim {m^2_{cr}\over T^2}$,
then the matter content
would be greater than the threshold and the singularity
could dissapear.

Any of the scalar fields $f_i$ can be expanded as
\begin{equation}
f(\tau,x)=\int {dk\over 2\pi}e^{ikx} [a_kf_k(\tau)+ a_{-k}^{\dagger}
f_k^*(\tau)],
\end{equation}
where  $a_k$ and $a_k^{\dagger}$ are the usual creation and annihilaton
operators. The modes $f_k$ satisfy
\begin{equation}
\ddot f_k+[k^2+\xi R e^{2\rho}]f_k=0\label{kg}
\end{equation}

In terms of the {\it functions} $\alpha(\tau)$ and $\beta(\tau)$
defined
through \cite{starzel}
\begin{eqnarray}
f_k &=& (2w_k)^{-1/2}[\alpha_k e^{-i\int w_k d\tau}
+\beta_k e^{i\int w_k d\tau}]\nonumber\\
\dot f_k &=& -i(w_k/2)^{-1/2}[\alpha_k e^{-i\int w_k d\tau}
-\beta_k e^{i\int w_k d\tau}]\nonumber\\
w_k^2&=& k^2 +\xi R e^{2\rho}\label{staro}
\end{eqnarray}
Eq.\ref{kg} reads
\begin{eqnarray}
\dot\alpha_k&=&{\beta_k\dot w_k\over2w_k}e^{2i\int w_k d\tau}\nonumber\\
\dot\beta_k&=&{\alpha_k\dot w_k\over2w_k}e^{-2i\int w_k d\tau}
\label{staro2}
\end{eqnarray}
with the normalization
condition $\vert\alpha^2_k\vert -\vert\beta_k\vert^2
=1$. In what follows we will assume that $ \vert\alpha_k\vert\simeq 1,
\vert\beta_k\vert\ll 1$. Therefore, the equation for $\beta_k$
can be easily solved
\begin{equation}
\beta_k\simeq {1\over 2}\int_{-\infty}
^{\tau}d\tau{\dot w_k\over w_k}e^{-2i\int^{\tau} w_k(\tau ') d\tau '}
\,\,\, ,
\end{equation}
where we assumed that
the matter fields are in the {\it in}-vacuum
($\beta_k(-\infty)=0$).
The energy density of created particles at time $\tau$ is  given by
\begin{equation}
\epsilon (\tau) = 2 N\int_0^{\infty}
dk w_k \vert\beta_k(\tau)\vert^2\label{eps}
\end{equation}

{}From the above equations it is easy to find an upper
bound for $\epsilon(\tau)$. The coefficient $\beta_k
(\tau)$ satisfies \footnote{For simplicity we assume that
${\dot w_k\over w_k}>0$ in the region of integration.}
\begin{equation}
\vert\beta_k\vert \leq {1\over 2}\int_{-\infty}
^{\tau}d\tau {\dot w_k\over w_k}
={1\over 4}\log\vert 1+{2\xi\ddot\rho (\tau )\over k^2}\vert
\end{equation}
With this bound, one can estimate the integral in Eq.\ref{eps}
by considering the cases $k^2\gg\vert\xi\ddot\rho (\tau )\vert $,
$k^2\sim\vert\xi\ddot\rho (\tau )\vert $,  and
$k^2\gg\vert\xi\ddot\rho (\tau )\vert $. The result is
\begin{equation}
\epsilon (\tau)\leq \alpha N\vert\xi\ddot\rho(\tau )\vert
\label{bound}
\end{equation}
where $\alpha$ is a number of order one. We will
set $\alpha =1$ in what follows.

{}From the solutions of Section II we find (see Eq.\ref{sol2})
\begin{equation}
\ddot \rho=-{1\over\kappa}\ddot A= - {1\over 2\kappa}
{\delta^2\over T^2}{A'^2-2A''A\over (A-{A'^2\over 4\kappa})^2}
\end{equation}
The couplings   $A(\phi)=e^{-\gamma\phi}$ and $\phi^{2n}$
satisfy that
$A'^2\sim A''A$  in the
weak coupling region. As a consequence,
\begin{equation}
\epsilon (\tau)\leq N\vert\xi\ddot\rho(\tau )\vert
\sim{N\kappa\vert\xi\vert\delta^2\over
T^2} g_c^2(\tau )
\end{equation}
where $g_c\sim g_{eff}$ is the $\phi$-dependent coupling
constant.
Therefore the energy in created particles vanishes near
the singularity. This means that the analysis of the
existence of singularities done in the previous
Section is valid and the solutions can be trusted
as long as  $\epsilon (\tau)\ll {m^2_{cr}\over T^2}$.

At this point one would conclude that particle creation
does not help to smear the singularities. We will now argue
that this
is not always the case. We have chosen the coordinate
$\tau$ in such a way that the singularity is at $\tau =-\infty$.
Moreover, we assumed that the quantum state of the matter
fields is the {\it in}-vacuum. This is the reason why
$\epsilon(\tau)$ vanishes on the singularity.

However, the semiclassical
equations for $\xi =0$ are time-reversal invariant.
As a consequence,
there are solutions where the singularity is located at
$\tau = +\infty$. If we choose again $\vert 0_{in}>$ as quantum
state,  the singularity will survive only if,
during the whole evolution, the energy
in created particles does not exceed the critical value.
Whether this condition is satisfied or not depends on
the different parameters ($\xi , N, \delta $)
and on the coupling
$A(\phi )$.
In any case, we see that
if there is enough particle creation  the proof of the existence
of singularities could be fundamentally flawed.

\section{CONCLUSIONS}

To summarize, in  the generic dilaton-gravity theory
considered here, when the matter content is below
the threshold,
the quantum effects of conformal matter fields produce weak
coupling singularities as long as $A'^2\gg\kappa A$ in the
weak coupling region.

In the particular case $\lambda^2=0$,
and when the matter content equals the critical value, the corresponding
critical
solution is the Milne universe. There is no solution for
$m^2>m^2_{cr}$.
On the other hand, when $\lambda^2\neq 0$,
the solutions for $m^2\geq m^2_{cr}$
do exist,
but they have no weak coupling singularities.

A similar critical behaviour
at the onset of black hole formation  has been recently
discovered numerically in the S-wave sector of four
dimensional general relativity, and analytically
in the RST model \cite{strtor}.
Our `cosmological' results suggest that,
also in the black hole problem,
the threshold should not depend on the coupling
constants of the theory, while the scaling should
depend.

Turning back to the cosmological singularities,
the situation is  different for non-conformal fields.
In this case, if the quantum state of the matter fields is
chosen to be the vacuum state near the singularity,
then the  singularity survives. However, for other quantum states,
if the energy in created particles exceeds the critical
value, the singularities  could
dissappear. Therefore, the semiclassical cosmological
singularities seem to be generic only when conformal
invariance is not broken.

Breaking of conformal invariance should  also
be important when analyzing the theory beyond the
semiclassical approximation. Indeed, most of the works
done in
$(1+1)$ dimensions - including this one -
assume, without justification, that
the spacetime metric is a classical object. This should be
justified in the full quantum theory.
An important
ingredient in the quantum to classical transition is
decoherence between macroscopic trajectories
\cite{varios}. It has been shown in
the context of $(3+1)$-dimensional
quantum cosmology that decoherence takes place if and
only if there is particle creation \cite{cmh}. This result
is independent of the number of dimensions. As a consequence,
a theory with conformal matter fields may  not have a
well defined classical limit!
Thus, in order to have a reasonable toy model
in $(1+1)$ dimensions, conformal invariance should be
broken.
It would be interesting to reanalyze the black-hole
puzzless in this context.

\section{ACKNOWLEDGMENTS}

We would like to thank Carmen N\'u\~ nez for useful discussions.
This research was supported by Universidad de Buenos Aires,
Consejo Nacional de Investigaciones Cient\'\i ficas y T\' ecnicas
and Fundaci\' on Antorchas.

\end{document}